\documentclass[aps,prl,twocolumn,groupedaddress,showpacs,color]{revtex4}
\usepackage{graphicx}

\begin{document}

\title{GEMMA experiment: three years of the search for the neutrino magnetic moment\\
{\sl\small (submitted to Physical Review Letters)}}

\author{A. G. Beda}
\author{E. V. Demidova}
\author{A. S. Starostin}
\email[]{Starostin@itep.ru}
\affiliation{ITEP (State Science Center, Institute for
  Theoretical and Experimental Physics, Moscow, Russia)}
\author{V. B. Brudanin}
\author{V. G. Egorov}
\email[{\rule[-5mm]{0mm}{10mm}}{\sl(Corresponding author): }]{egorov@nusun.jinr.ru}
\author{D. V. Medvedev}
\author{M. V. Shirchenko}
\author{Ts. Vylov}
\affiliation{JINR (Joint Institute for Nuclear Research, Dubna, Russia)}

\date{\today}

\begin{abstract}
 The result of the 3-year neutrino magnetic moment
measurement at the Kalinin Nuclear Power Plant ({\sl KNPP})
with the {\sl GEMMA} spectrometer is presented. Antineutrino-electron scattering is investigated. A high-purity
germanium detector of 1.5 kg placed at a distance of 13.9 m from the
3~GW$_{\rm th}$ reactor core is used in the spectrometer. The antineutrino
flux is 2.7$\times$10$^{\rm 13}\; \bar\nu_e$/cm$^2$/s. The
differential method is used to extract $\nu${\em -e}
electromagnetic scattering events. The scattered electron spectra
taken in 5184+6798 and 1853+1021 hours for the reactor {\sl ON} and {\sl
OFF} periods are compared. The upper limit for the neutrino
magnetic moment $\mu _{\nu } < $3.2$\times$10$^{\rm -11}\mu_B$ at
90{\%} CL is derived from the data processing.
\end{abstract}

\maketitle

\section{Introduction}

The Minimally Extended Standard Model predicts a very small magnetic moment for the massive neutrino ($\mu _\nu\sim10^{-20}\mu _B$) which cannot be observed in an experiment at present.
On the other hand, there is a number of extensions of
the theory beyond the Minimal Standard Model where the {\sl Majorana} neutrino magnetic moment (NMM) could be at the level of
$10^{-(10-12)}$~$\mu_B$
irrespective of the neutrino mass \cite{Voloshin86,Fukugita87,Pakvasa03,Gorchtein06,Bell06}.
At the same time, from general considerations \cite{Bell05,Bell06d} it follows that
the {\sl Dirac} NMM could not exceed $10^{-14}\mu_B$.
Therefore, observation of an NMM value higher than $10^{-14}$~$\mu_B$ would be evidence for New Physics and, in addition, indicate \cite{Kayser08,Giunti08,Studenikin08} undoubtedly that the neutrino is a Majorana particle.

It is rather important to make laboratory NMM measurements sensitive enough to reach the $\sim$10$^{-11}\mu_B$ region. The Savanna River experiment by Reines' group could be considered as the beginning of such measurements. Over a period of thirty years sensitivity of reactor experiments increased by a factor of three only -- from $(2-4)\times10^{-10}\mu_B$\cite{Reines76,Vogel89} to $(6-7)\times10^{-11}\mu_B$\cite{TEXONO06,GEMMA07}. Similar limits were obtained for solar neutrinos\cite{SK04,BOREXINO}, but due to 
the MSW effect (as well as matter-en\-han\-ced oscillations in the Sun)
their flavor composition changes and therefore the solar NMM results
could differ from the reactor ones.

In this paper, the results of the 3-year NMM measurement by the collaboration of
ITEP (Moscow) and JINR (Du\-b\-na) are presented.
The measurements are carried out with the {\sl GEMMA} spectro\-meter\cite{Beda98,Beda04,GEMMA07} at the 3~GW$_{\rm th}$ reactor of
the Kalinin Nuclear Power Plant (KNPP).

\section{Experimental approach}
A laboratory measurement of the NMM is based on its contribution to the $\nu$-$e$
scattering. For nonzero NMM the $\nu$-$e$ differential cross
section is given \cite{Vogel89} by a sum of the {\sl weak} interaction cross section
($d\sigma_W/dT$) and the {\sl electromagnetic} one ($d\sigma_{EM}/dT$):
\begin{figure}[hb]
 \setlength{\unitlength}{1mm}
  \begin{picture}(85,56)(0,0)
   \put(-1,-2){\includegraphics{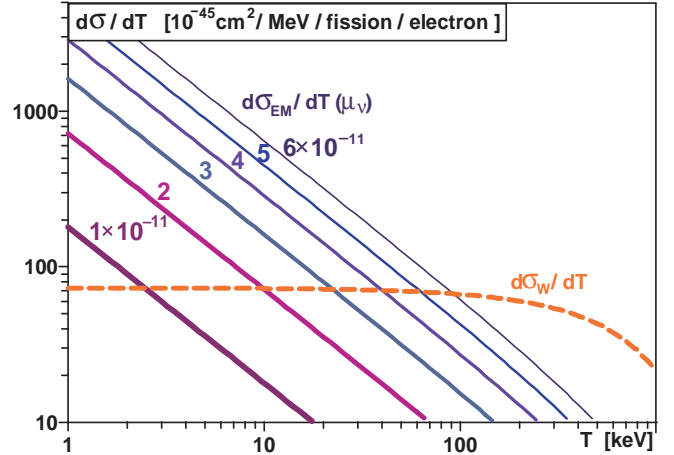}}
  \end{picture}
 \caption{Weak (W) and electromagnetic (EM) cross-sections calculated for several
 NMM values.}
 \label{Fig.W/EM}
\end{figure}

\begin{eqnarray}
\frac{d\sigma_W}{dT} = \frac{G_F^2 m_e }{2\pi}\left[ \left(1\!-
\frac{T}{E_\nu}\right)^{\!\!2}\!\left(1\!+2\sin^2\theta_W\right)^2 \!\!+  \right. \hspace{5mm} \nonumber\\[-2mm]
\left.  +  4\sin^2\theta_W -\! 2\left(1\!\!+2\sin^2\theta_W\right)\sin^2\theta_W
\frac{m_eT}{E_\nu^2}\right],
\label{eq.dsW/dT}
\end{eqnarray}
\vspace*{-3mm}
\begin{equation}
\frac{d\sigma_{EM}}{dT} =\pi r_0^2 \left( {\frac{\mu _\nu }{\mu _B }}
\right)^2\left( {\frac{1}{T} - \frac{1}{E_\nu }}
\right)\;\;,
\label{eq.dsEM/dT}
\end{equation}

where $E_\nu$ is the incident neutrino energy, $T$ is the electron recoil
energy, $\theta_W$ is the Weinberg angle and $r_0$ is the electron radius ($\pi r_0^2=2.495\times10^{-25}$ cm$^2$).

Figure~\ref{Fig.W/EM} shows differential cross sections (\ref{eq.dsW/dT})
and (\ref{eq.dsEM/dT}) averaged over the typical antineutrino reactor spectrum
vs the electron recoil energy.
One can see that at low recoil energy ($T\ll E_\nu$) the value of $d\sigma_W/dT$ becomes
almost constant, while $d\sigma_{EM}/dT$ increases as $T^{-1}$, so that the
lowering of the detector threshold leads to a considerable increase in
the NMM effect with respect to the weak unremovable contribution.

To realize this useful feature in our GEMMA spec\-tro\-meter\cite{GEMMA07}, we use a 1.5 kg HPGe detector with the energy threshold as low as 3.0 keV. To be sure that there is no efficiency cut at this energy, the "hard" trigger threshold was twice lower (1.5 keV).

Background is suppressed in several steps. First, the detector is placed inside a cup-like NaI crystal with 14 cm thick walls surrounded with 5 cm of electrolytic copper and 15 cm of lead. This active and passive shielding reduces external $\gamma$-background in the ROI to the level of $\sim2$~counts/keV/kg/day. Being located just under reactor \#2 of the KNPP (at a distance of 13.9 m from the reactor core, which corresponds to the antineutrino flux of $2.7\times 10^{13}\;{\rm \bar\nu_e/cm^2/s}$), detector is well shielded against the hadronic component of cosmic rays by the reactor body and technologic equipment (overburden$\simeq$70 m w.e.). The muon component is also reduced by a factor of $\sim$10 at $\pm20^\circ$ with respect to the vertical and $\sim$3 at $70^\circ\!-80^\circ$, but a part of residual muons are captured in massive shielding and thus produce neutrons which scatter elastically in Ge and give rise to a low-energy background. To suppress it, the spectrometer is covered with additional plastic scintillator plates which produce relatively long $\mu$-veto signals. Special care is taken to reduce non-physical low-amplitude circuit noise (afterpulses, radio frequency interference, microphonism, etc.). In particular, the detector signal is processed by three parallel independent electronic channels with different shaping time, which allows performing a primitive Fourier analysis \cite{Garcia92} \`{a} posteriori and thus discriminating artefact signals (Fig.\ref{Fig.Matrix}).

\section{Data taking and
pro\-ces\-sing}

In order to get a recoil electron spectrum, we use a differential
method comparing the spectra measured at the reactor operation (ON)
and shut down (OFF) periods. In our previous work we considered Phase-I (13 months'
measurement from 08.2005 to 09.2006, including 5184 and 1853
ho\-urs of the reactor ON and OFF periods, respectively). Today we can add Phase-II -- 19 months from 09.2006 to 05.2008.
Unfortunately, for some organizational and technical reasons, there were several long interrupts in the measurement. After preliminary selection, 6798 ON-hours and 1021 OFF-hours of live time were found to be available for analysis.

During the measurements, the signals of the HPGe detector,
anticompton NaI shielding and outer anticosmic plastic counters, as well as dead-time information, are collected on an event-by-event basis.
Detection efficiency just above the threshold is checked with a pulser.
The neutrino flux monitoring in the ON period is carried on via the reactor
thermal power measured with accuracy of 0.7\%.

The collected data are processed in several steps. First, we reject those files which correspond to the periods of liquid nitrogen filling and any mechanical or electrical work at the detector site, as they could produce a noise. Second, we analyze energy spectra produced for each hour in order to check stability of $\gamma$-background. If any visible excess of 81~keV ($^{133}$Xe), 250~keV ($^{135}$Xe) or 1294~keV ($^{41}$Ar) $\gamma$-line occurs, the files are removed.
Third, the level of low-amplitude non-physical noise is checked second by second, and those seconds which contain more than 5 events with $E>$2~keV are rejected.
Fourth, we reject those events which are separated by a time interval shorter than 80~ms or equal to $n\cdot20.0\pm0.1$~ms (in such a way we suppress the noise caused by mechanical vibrations and 50~Hz power-line frequency).
\begin{figure}[hb]
 \setlength{\unitlength}{1mm}
  \begin{picture}(85,80)(0,0)
   \put(0,-3){\includegraphics{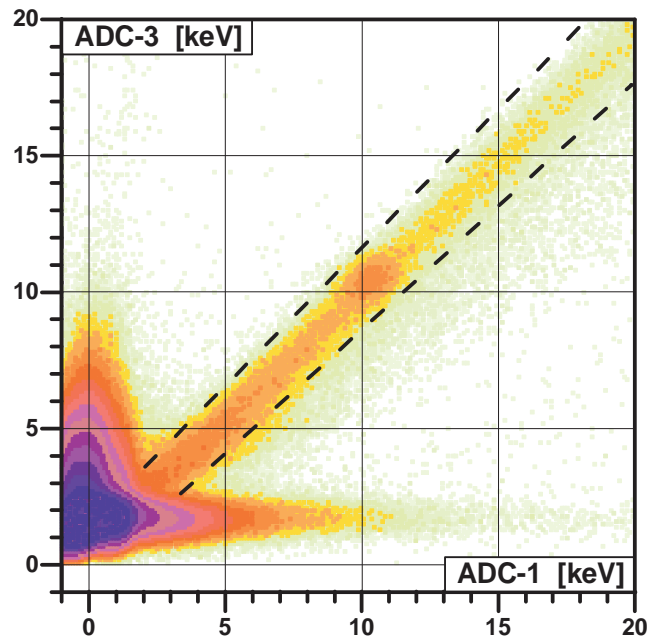}}
  \end{picture}
 \caption{(Color online) Example of the primitive Fourier analysis done with two different shaping-times: ADC-1 operates with 2~$\mu$s pulses, and ADC-3 operates with 12~$\mu$s pulses. (Color intensity scale is logarithmic).}
 \label{Fig.Matrix}
\end{figure}

Then, we build three plots similar to that shown in Fig.~\ref{Fig.Matrix} and select only those events which fall (within the energy resolution) into diagonals, thus rejecting low- and high-frequency noise. As a result, we obtain energy spectra for the ON and OFF periods which must be normalized by the corresponding active time. Since the described selection of events is complicated, it is difficult to count active time in a proper way. To avoid possible errors caused by this procedure, both the ON and OFF spectra are normalized by the intensity of the background $\gamma$-lines which are definitely known in time. These are the 1173~keV and 1333~keV lines of $^{60}$Co, the 1461~keV line of $^{40}$K and a less reliable 238~keV line of $^{212}$Pb. The above radiation originates from the pollution of the internal parts of the spectrometer and therefore must be independent of the reactor operation.

\begin{figure}[ht]
 \setlength{\unitlength}{1mm}
  \begin{picture}(85,40)(0,0)
   \put(0,-3){\includegraphics{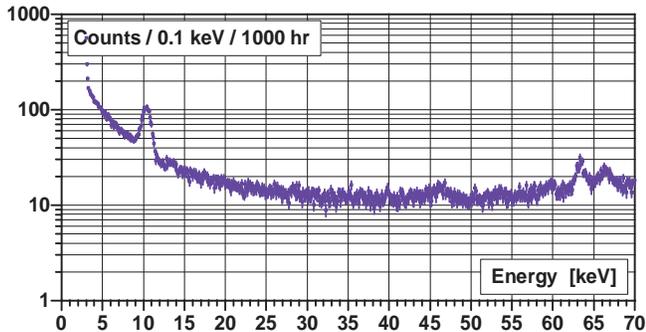}}
  \end{picture}
 \caption{Fragment of the experimental ON spectrum.}
 \label{Fig.ON-spk}
\end{figure}

\begin{figure}[th]
 \setlength{\unitlength}{1mm}
  \begin{picture}(85,40)(0,0)
   \put(1,-3){\includegraphics{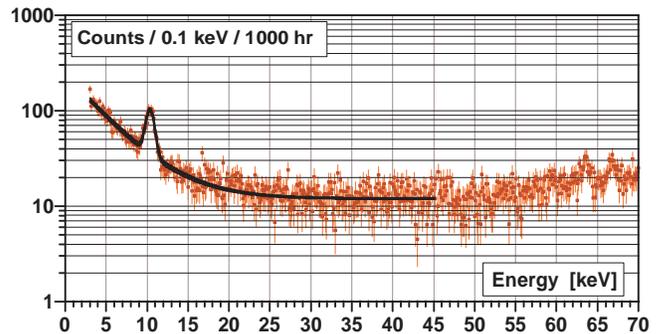}}
  \end{picture}
 \caption{Fragment of the experimental OFF spectrum.}
 \label{Fig.OFF-fit}
\end{figure}

To extract the $\mu_\nu$ value from the normalized ON and OFF spectra, we use two procedures. One of them was described in detail in our previous work \cite{GEMMA07}. It consists in the channel-by-channel comparison of the spectra (taking into account the Weak contribution) and then averaging of the extracted $X_i$ values over the ROI. Here $i$ is the 0.1 keV-channel number, and $X$ stands for an NMM squared in terms of 10$^{-10}$ Bohr magnetons:
\begin{equation}\label{eq.X}
X\equiv \left(\frac{\mu_\nu}{10^{-10}\mu_B}\right)^2 \;.
\end{equation}
The above procedure is perfectly reliable and does not depend on the background structure. Unfortunately, the ON and OFF periods are not equal from the point of view of statistics (compare error bars in Figs.~\ref{Fig.ON-spk} and \ref{Fig.OFF-fit}). A usual OFF period is much shorter, and, therefore, the final sensitivity is limited by the background uncertainties. On the other hand, today, after three years of data taking, we know the ROI background structure with more confidence. It gives us the right to introduce additional information in our analysis, namely, to state that our background is {\sl a smooth curve}.

To implement this idea, we fit the background in the ROI from 2.9 keV to 45 keV with a parametrized smooth function (an example of such fit with a sum of Gaussian, exponential and linear functions is shown in Fig.~\ref{Fig.OFF-fit}; other functions produce slightly different results, the systematic error includes their spread). Then we compare the ON spectrum channel by channel with the obtained curve (to be more precise, with a narrow corridor of the width given by the fitting uncertainty). Applying this advanced procedure to the total statistics of Phases I+II, we get the following NMM limit:
\begin{equation}
\mu_\nu<3.2\times10^{-11}\mu_{\rm B}\hspace{5mm}{\rm (90\%CL)}
\end{equation}

\section{Conclusion}

The experimental NMM search with the GEMMA spectrometer has been going on at the
Kalinin Nuclear Power Plant (Russia) since 2005. The HPGe detector of 1.5 kg placed  13.9~m under the core of the 3~GW$_{\rm th}$ water-moderated reactor is exposed to the antineutrino flux of $2.7\!\times\!10^{13}\bar\nu_e/{\rm cm}^2/{\rm s}$. As a result of the 3-year measurement (about 13000 ON-hours and 3000 OFF-hours of live time), the upper limit of 3.2$\times$10$^{-11}\mu_{\rm B}$ at 90\%CL was found for the NMM.

At present, the data taking is in progress, but analysis of the data indicates that the sensitivity limit of the setup is almost reached. To improve it, we prepare significant upgrading of the spectrometer (GEMMA-2). Within the framework of this project we plan to use the antineutrino flux of $\sim5.4\times10^{13} \bar\nu_e /{\rm cm}^2/{\rm s}$, increase the
mass of the germanium detector by a factor of four and decrease the level of the background. These measures will provide the possibility of achieving the NMM limit at the level of $1.5\times10^{-11}\mu_{\rm B}$.

\begin{acknowledgments}
The authors are grateful to the directorates of ITEP and JINR for constant support of this work and especially to M.V.~Danilov for his important comments. The authors appreciate the administration of the KNPP and the staff
of the KNPP Radiation Safety Department for permanent assistance in the
experiment.
This work is supported by the Russian State Corporation ROSATOM
and by the Russian Foundation for Basic Research, pro\-ject 09-02-00449.
\end{acknowledgments}

\bibliography{GEMMA}
\end{document}